\documentclass{optica-article}

\journal{opticajournal} 

\articletype{Research Article}

\usepackage{lineno}
\usepackage{siunitx}
\DeclareSIUnit{\decibelm}{dBm} 

\begin{document}
\title{Ising accelerator with a reconfigurable interferometric photonic processor}

\author{José Roberto Rausell-Campo,\authormark{1,*,$\dag$} Nayem Al Kayed,\authormark{2,*,$\dag$}, Daniel Pérez-López\authormark{3}, A. Aadhi\authormark{2,4}, Bhavin J. Shastri\authormark{2}, José Capmany Francoy\authormark{1}}

\address{\authormark{1}Photonics Research Lab, iTEAM. Universitat Politècnica de Valencia, Camino de Vera, s/n 46022 Valencia, Spain\\
\authormark{2}Centre for Nanophotonics, Department of Physics, Engineering Physics and Astronomy, Queen’s University, Kingston, ON K7L 3N6, Canada\\
\authormark{3}IPronics Programmable Photonics S.L, Av. de Blasco Ibáñez, 25, Valencia, Spain\\
\authormark{4} Optics and Photonics Centre, Indian Institute of Technology Delhi, Delhi 110016,
India. \\
\authormark{$\dag$}These authors contributed equally to this work.}

\email{\authormark{*}joraucam@upv.es, nayemal.kayed@queensu.ca} 

%


\begin{abstract*} 
 The general-purpose programmable photonic processors offer a scalable and reconfigurable solution for a wide range of RF and optical applications. Therefore, implementing photonic Ising machines using programmable processors leverages the advantages of high speed and parallelism, enabling efficient hardware acceleration for finding ground-state solutions to combinatorial optimization problems. In this work, we demonstrate a novel programmable photonic Ising solver based on a hexagonal mesh general-purpose programmable photonic platform. The integrated system allows reconfigurable matrix multiplication and computes the Hamiltonian iteratively using an annealing algorithm that facilitates spin updates and effectively searches for the ground state. As a proof of concept, we experimentally solve two benchmark optimization problems, a fundamental three-node ferromagnetic coupling problem with external bias that demonstrates nontrivial spin interactions, and a four-node Max-Cut problem with arbitrary coupling matrices. Furthermore, to establish a large-scale capability, we emulated Ising problems with sizes up to N = 50, achieving success probabilities exceeding 80\%. Additionally, we examined the impact of errors, such as phase and coupling, on the performance of the programmable photonic Ising machine. Our general-purpose photonic Ising machine paves the way for implementing large-scale, programmable architectures for solving optimization problems.

\end{abstract*}

\section{Introductions} 
Combinatorial optimization problems are among the most challenging computation, yet they underlie numerous real-world tasks, ranging from logistics and finance to circuit design, communications and drug discovery \cite{lucas2014ising, martello1990knapsack, glover2003handbook}. As problem sizes increase, traditional von Neumann processors break down due to an increased complexity and exponentially large search space, often demanding prohibitively large power consumption and computational resources \cite{waldrop2016chips,trinh2020high}. As a result, the demand for more specialized, energy-efficient hardware accelerators for large-scale optimization has intensified in recent years, due to increased computational need and the slowing of Moore's law \cite{waldrop2016chips}. Ising model is one of the compelling frameworks for tackling such large-scale optimization problems \cite{ising1925beitrag}, wherein binary spins $(\sigma_i = \pm 1)$ and the interaction matrix $J_{ij}$ represent the defined problems. The energy or Hamiltonian of the system is expressed by
\begin{equation}
    H = \frac{1}{2}\sum_{<i,j>}J_{ij}\sigma_i\sigma_j - \sum_{i} h\sigma_i
    \label{eq:sing}
\end{equation}
where $J_{ij}$ denotes interaction strengths between spins $\sigma_i$ and $\sigma_j$, and $h_i$ is an external field acting on spin $\sigma_i$. A broad class of computationally intractable problems can be mapped onto an Ising Hamiltonian with specific spin couplings, where the ground state represents the optimal solution. In the Ising model, finding the ground state of the system by minimizing its energy is equivalent to solving the associated combinatorial optimization problem \cite{lucas2014ising, kochenberger2014unconstrained}. Motivated by the computational efficiency of the Ising model, a variety of Ising machine architectures have been developed, each leveraging the unique advantages of different physical substrates including, CMOS-based annealers \cite{yamaoka2015ising}, magnetic tunnel junctions \cite{okuyama2015magnetic}, superconducting quantum devices \cite{lanting2014entanglement}, and photonics \cite{mcmahon2016fully,inagaki2016large,pierangeli2019large, Prabhu2020, gao2024} to explore the energy landscape more efficiently than conventional digital processors.   Among all Ising machine architectures, photonic implementations are emerging prominently due to the inherent parallelism, ultrafast operation, high coherence, and room-temperature operation of photonic architectures \cite{hamerly2019experimental, bueno2022reinforcement}. Previous demonstrations using bulk optics and fiber-based approaches have shown the feasibility of large-scale optical hardware accelerators based on iterative matrix-vector multiplications \cite{inagaki2016large, mcmahon2016fully, honjo2021100}. However, various factors such as precise alignment, instability, and large physical footprint have driven research toward a compact integrated photonic platform \cite{burmistrov2023optically}. 

Recently, general-purpose programmable photonic platform based on programmable chip built using a hexagonal waveguide mesh, where each side of the hexagon contains two waveguides controlled by tunable Mach–Zehnder Interferometer (MZI) units, have emerged as novel photonic computing architectures \cite{perez2017multipurpose}. By adjusting phase shifters in these MZIs, the photonic chip can be reprogrammed to perform a wide range of optical processing functions, including filtering, ring cavities, complex Coupled Resonator Optical Waveguide (CROW) devices, delay lines, arbitrary waveform generators and linear optics transformations (such as a CNOT gate) \cite{Perez-Lopez2024, rausell}. A significant strength of this platform lies in its self-configuration capabilities, which employ optimization algorithms, like genetic algorithms and particle swarm optimization to automatically adapt to imperfections, including uneven losses and phase offsets \cite{Perez-Lopez2020}. However, the use of programmable silicon photonics for solving combinatorial optimization problems, enabling reconfigurability and the ability to address a wide range of optimization tasks through the Ising model, has not yet been explored. Combining the advantages of photonics with recent advancements in general-purpose programmable platforms could enable large-scale integration and significantly reduce production costs. Here, we present the first experimental demonstration of an on-chip programmable photonic Ising machine that efficiently performs eigenvalue decomposition and implements an annealing algorithm to solve Ising problems. As a proof of concept, our demonstration uses the Smartlight programmable photonic processor based on a hexagonal mesh architecture with reconfigurable MZIs \cite{Perez-Lopez2024}. In this system, spin variables are encoded in optical phases, and the Ising Hamiltonian is recovered through a single intensity measurement \cite{wu2022photonic}. A digital feedback loop iteratively updates the spin configuration until a low-energy state is reached \cite{kirkpatrick1983optimization, farhi2014quantum}.Our benchmarking demonstrations on a three-node problem and a four-node Max-Cut problem show success probabilities approaching unity in most instances, even after only a few iterations, highlighting the robustness of our programmable photonic Ising machine. Furthermore, we investigate the scalability of our architecture (up to N = 50), as well as the impact of noise, latency, energy efficiency, and error mitigation strategies, to assess its viability for large-scale implementation. This work paves the way for fully integrated, programmable photonic architectures capable of addressing a broad class of combinatorial optimization problems at unprecedented speeds and energy efficiencies.

\begin{figure}[h]
    \centering
    \includegraphics[width=1.0\textwidth]{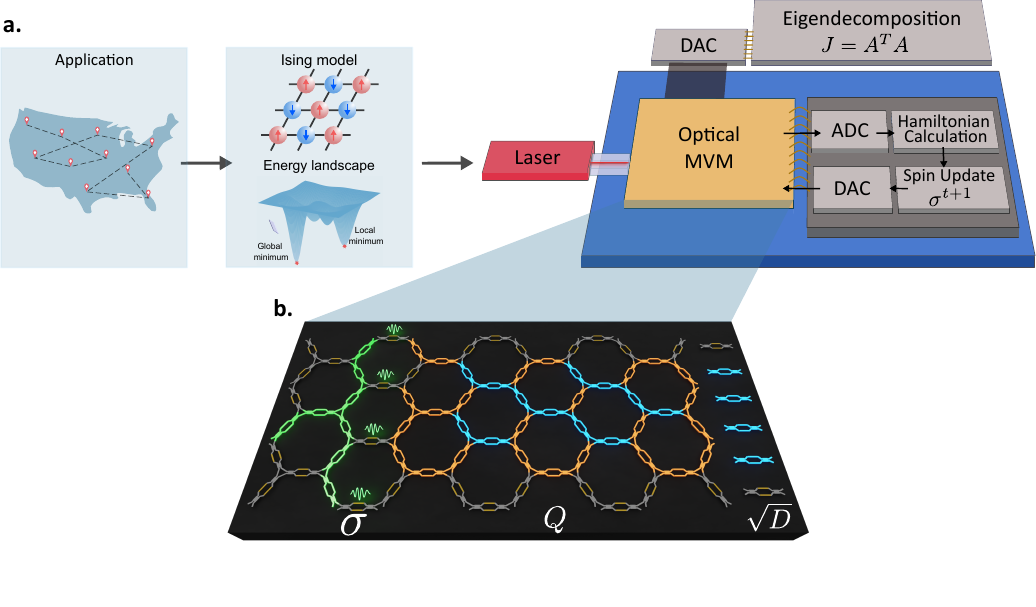} 
    \caption{\textbf{a.} An example of a combinatorial optimization problem, the Ising model mapping with its energy landscape and the schematic of a photonic-electronic integrated-based Ising solver. The Ising solver works as a phase encoding and intensity detection annealer where the coupling matrix $J$ is decomposed into its eigenvectors and eigenvalues $J = Q^{T}DQ$. The product $A = \sqrt{D}Q$ is encoded in the optical matrix-vector multiplier (MVM) with the spin state $\sigma$. After photo-detection, the Hamiltonian is calculated, and the new spin state is obtained. The spin state of the optical MVM is updated, and the process is repeated during $N$ iterations. \textbf{b} Representation of the general-purpose photonic processor with a hexagonal topology. The green path represents a splitter tree used to divide light among the input paths where spin encoding is carried out. The orthogonal matrix $Q$ is implemented using the rectangular (Clements) configuration for unitary matrix multiplication, where the MZIs in orange are in a fixed state while the blue MZIs are tunable. Finally, an ancillary array of MZIs is used to encode the diagonal matrix $D$.}
    \label{fig:conceptual}
\end{figure}

\section{Working Principle and Architecture}
The use of a programmable photonic Ising solver to address combinatorial optimization problems leverages matrix-vector multiplications, an algorithm for energy calculation, and spin updates. Spin encoding is achieved using an array of dual-drive MZIs, where each MZI tunes its phase to represent the spin vector. Although arbitrary matrix-vector multiplications cannot be directly implemented using programmable photonics for Hamiltonian computation, decomposing the coupling matrix into its eigenvalue decomposition enables optical encoding. As a result, unlike conventional architectures \cite{prabhu2020accelerating, wu2025monolithically}, programmable photonics allow for unitary matrix multiplication directly in the optical domain, making them well-suited for Hamiltonian computation \cite{rausell}. Detailed descriptions of the spin encoding and matrix implementation are provided in Supplementary Section I. To find the ground-state solution, an annealing process iteratively computes the Ising energy and updates the spin configuration to minimize the energy. Our algorithm employs a probabilistic update rule: spin flips are always accepted if the change in the Hamiltonian ($\Delta$H) is equal to or less than zero; otherwise, spins are updated with a certain probability. The iterative process stops after a predefined maximum number of iterations, and the spin configuration with the lowest observed Hamiltonian is selected as the solution. This approach enables a broad exploration of the energy landscape, fast convergence and high accuracy toward an optimal solution. A detailed implementation of the algorithm is provided in the Methods section. The programmable processor enables precise encoding of interaction matrices and spin configurations, leveraging the inherent parallelism and high-speed operation of photonic computing. 

The programmable photonic processor, based on a hexagonal mesh architecture, has been previously demonstrated in a variety of applications \cite{Perez2017, Perez-Lopez2024}. Here, we implement the photonic Ising solver using the same platform, leveraging its capability to perform reconfigurable matrix-vector multiplications in the optical domain and spin updates in the electronic domain. Figure \ref{fig:conceptual}\textbf{a} illustrates the schematic implementation of the programmable photonic Ising solver. In this architecture, a single Hamiltonian calculation involves the multiplication of three matrices, $\sqrt{D}Q\sigma$, which is carried out in the optical domain using the general-purpose processor as an MVM accelerator, while the spin update is performed in the electronic domain. An schematic of the fabricated photonic chip in the Smartlight system is shown in Fig.\ref{fig:conceptual}\textbf{b}. The processor is an integration of 17 hexagonal cells, comprising a total of 72 programmable unit cells (PUCs). Technical specifications and additional details are provided in the Methods section. The spin encoding within the hexagonal mesh grid begins with a splitter tree that divides the incoming light into multiple paths. As shown in the figure, the four input sections can be programmed to encode four spin values, either in amplitude or phase. In our implementation, spin values are encoded using phase shifts of 0 and $\pi$ achieved via three sets of dual-drive MZIs. Similarly, the array of MZIs implements the reconfigurable interaction matrix via $Q$ and the diagonal matrix $\sqrt{D}$. The matrix $Q$, which contains the eigenvectors of the coupling matrix $J$, is realized using an array of MZIs arranged in either a triangular \cite{Reck1994} or rectangular (Clements) configuration \cite{Clements2016}. The optical output intensities from the photonic processor are measured using photodetectors. These measured intensities are then used to compute the energy of the system $H(\sigma)$ using a CPU, which updates the spin configuration based on the computed energy (see Methods). The proposed architecture enables the most computationally intensive part of the Hamiltonian solver, namely, the eigendecomposition, to be offloaded to a photonic integrated platform, leveraging the inherent speed and parallelism of photonic computation. Unlike previous architectures, we implement both the unitary matrix and the diagonal matrix directly in the photonic domain. Notably, while the general-purpose programmable processor is used to implement unitary matrix multiplications, the inclusion of an ancillary array of MZIs in the architecture enables the realization of symmetric non-unitary matrices as well, thereby solving a broader class of problems, especially those involving dissipation, loss, or real-valued transforms. An image of the fabricated general-purpose system is presented in Fig.\ref{fig:operating}\textbf{a}.

\section{Experimental Results}
\begin{figure}[h]
    \centering
    \includegraphics[width=1.0\textwidth]{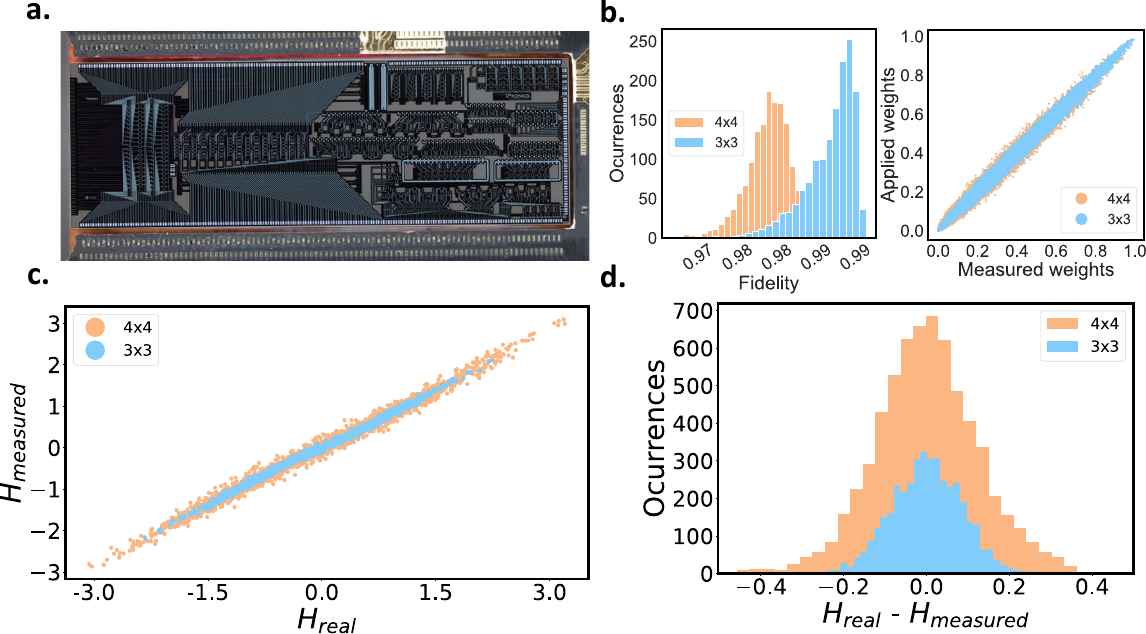} 
    \caption{\textbf{a.} Image of the programmable photonic chip in the Smartlight processor. \textbf{b} Experimental fidelity and weight distribution of 1500 random unitary matrices with 3x3 (blue) and 4x4 (orange) sizes. \textbf{c.} Comparison between the real and measured Hamiltonian for different random coupling matrices and spin configurations \textbf{d.} Difference between the real and measured Hamiltonian for different random coupling matrices and spin configurations.}
    \label{fig:operating}
\end{figure}

The calculation of the Hamiltonian involves the implementation of on-chip vector-matrix multiplication, as described in the Methods section. The spin vector and diagonal matrix are encoded using an array of MZIs configured to represent the amplitude terms. For the spin vector, where all elements have an amplitude of 1, the MZIs are set to the \textit{bar} state (zero coupling). The sign of the spin vector components is incorporated by simultaneously tuning both phase shifters of each MZI. A detailed description of the encoding process for vector elements using dual-drive MZIs is provided in Section I of the supplementary material. The elements of the diagonal matrix are normalized to values between 0 and 1 by dividing the array by its maximum value. The normalization factor is stored, and its square is subsequently used to scale the measured photocurrents. The eigenvector matrix $Q$ is encoded within a rectangular photonic architecture for unitary transformations. The Smartlight processor enables the encoding of unitary matrices of sizes $3 \times 3$ and $4 \times 4$ with fidelities of $99.2 \pm 0.3$ and $98.4 \pm 0.3$, respectively, and bit precisions exceeding five bits. Figure \ref{fig:operating}\textbf{b} illustrates the fidelity and weight distribution of 1500 sampled matrices for these two sizes. To evaluate the performance of the programmable mesh for Hamiltonian calculation, we tested 500 and 400 random coupling matrices of sizes $3 \times 3$ and $4 \times 4$, respectively. Each coupling matrix was decomposed to encode the $Q$ and $\sqrt{D}$ matrices. Subsequently, all possible spin configurations were generated, $2^3$ for the $3 \times 3$ case and $2^4$ for the $4 \times 4$ case, and the Hamiltonian was computed for each configuration using (\ref{eq:H_photodetection}). This resulted in 4000 and 6400 Hamiltonians for the $3 \times 3$ and $4 \times 4$ cases, respectively. A comparison between the theoretical Hamiltonian and the measured Hamiltonian for each spin configuration is shown in Fig. \ref{fig:operating}\textbf{c}, achieving an $r^2$ value of 0.99 for the $3 \times 3$ case and 0.98 for the $4 \times 4$ case. The error, defined as the difference between the theoretical and measured values, is presented in Fig. \ref{fig:operating}\textbf{d}, with mean squared errors (MSE) of 0.007 and 0.016, respectively.

\begin{figure}[ht]
    \centering
    \includegraphics[width=1.0\textwidth]{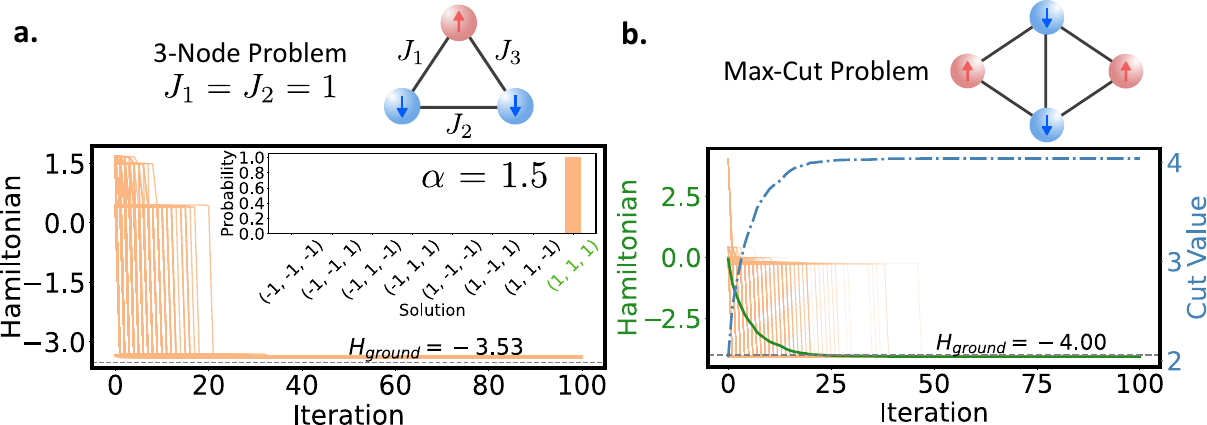} 
    \caption{Example of the evolution of the Hamiltonian (orange lines) during the annealing algorithm for 1000 models and 100 iterations. The dotted horizontal line represents the ground state of the problem. Results are presented for \textbf{a.} 3-Node problem with $J_{1} = J_{2} = 1$ and $\alpha = \frac{J_{3}}{J_{1}}= 1.5$. Inset graph represents the probability distribution of the solutions for the 1000 models and shows in green the optimal solution. \textbf{b.} Max-Cut problem with 4 nodes. The green line represents the mean of Hamiltonian evolution and the blue dotted line the evolution of the cut value.}
    \label{fig:results}
\end{figure}
Using the photonic processor, we addressed the three-node spin interaction and the four-node MaxCut problems \cite{marandi2014network}. The three-node problem represents the simplest model with nearest-neighbour antiferromagnetic interaction while still preserving symmetry and key properties of lattice systems, such as disorder and phase transitions. For the demonstration of three-node nearest-neighbour interactions, the coupling matrix and the system parameter are defined as $J_1 = J_2 = 1$ and $\alpha = \frac{J_3}{J_2}$, respectively. The goal of the Ising problem is to determine the most stable spin configuration, corresponding to the minimum energy state. The problem was mapped for $\alpha$ values ranging from -1.5 to 1.5 in increments of 0.1, with $h = 0.01$. For each value of $\alpha$, the Ising model was run for 100 iterations, each with 1000 random initial conditions. The final solutions and the probability of reaching each solution were recorded. The evolution of the Hamiltonian over the iterations is shown in Fig. \ref{fig:results}\textbf{a}. The inset depicts the probability distribution of solutions for $\alpha = 1.5$, where the optimal solution is highlighted in green, corresponding to a minimum energy value of -3.53. The ground state was successfully reached within 100 iterations for all initial conditions. Results for the remaining trained models are provided in Section II of the Supplementary Material. We observed a maximum success probability of unity for almost all $\alpha$ values, except for $\alpha = -0.9$, $-1.0$, $-1.3$, and $-1.5$. For these values, the system failed to overcome the local minima to reach the global minimum. However, introducing noise or employing algorithmic modifications can improve the performance of the system \cite{kalinin2025, kayed2025}. 

By reconfiguring the programmable photonic processor, we further efficiently found the optimal solutions for a common benchmarking problem, the Max-Cut problem \cite{Commander2009}. Max-Cut is a classical combinatorial optimization problem that involves partitioning the nodes of a graph into two subsets such that the sum of the weights of the edges connecting the two subsets is maximized. In our experiment, the reconfigurability of the photonic processor enabled us to implement four 4-node Ising problems with five different coupling matrices. Using an array of MZIs, we mapped the Max-Cut weights onto the Ising model. For each coupling matrix, the system was initialized with 1000 random initial conditions and run over 100 iterations. Figure \ref{fig:results}\textbf{b} also depicts the evolution of the corresponding Max-Cut value (blue dotted line) over the iterations. The cut value was calculated using the relation $ \text{Cut Value} = \frac{1}{2}\sum_{ij}J_{ij} (1 - \sigma_i \sigma_j) $. Our programmable photonic Ising architecture reaches the success probability of 1.0 across all initial configurations. Additional results are provided in Section III of the Supplementary Material. Similar to the three-node problem, we obtain the success probability of 1 for all the measured problems.

\section{Discussion}

\begin{figure}[h]
    \centering
    \includegraphics[width=1.0\textwidth]{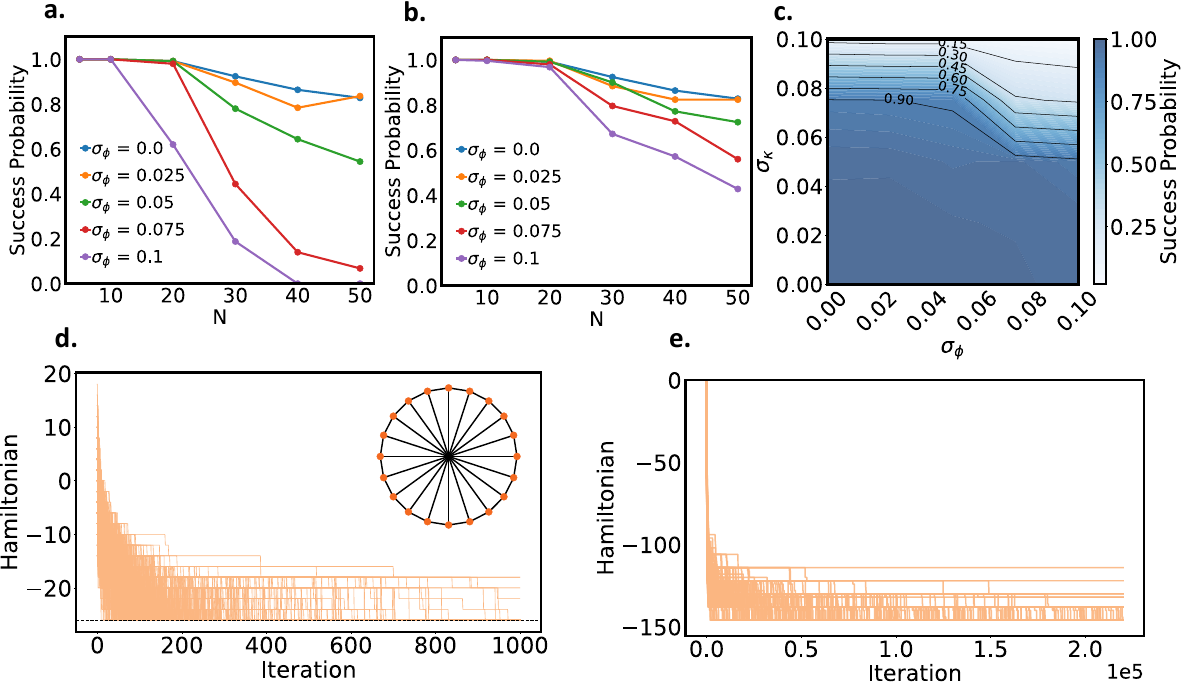} 
    \caption{Performance simulations of the programmable Ising machine: \textbf{a.} Max-Cut with varying phase errors and problem sizes,  \textbf{b.} Max-Cut with varying coupling errors and problem sizes,  \textbf{c.} Möbius ladder with $N=20$ under different phase and coupling errors, \textbf{d.} Hamiltonian evolution for the ideal case of the Möbius ladder with $N=20$, and \textbf{e.} Hamiltonian evolution for the Möbius ladder with $N=100$.}

    \label{fig:simulation}
\end{figure}

The implementation of large-scale matrix-vector multiplications is critical for real-world applications and unlocking the full potential of analog photonic Ising machines. Therefore, we analyzed several key aspects related to the scalability and precision of our photonic matrix multipliers, including insertion losses, phase errors, and energy efficiency of the large-scale programmable photonic Ising architecture.
To demonstrate the scalability of the architecture, we emulated the system with problem sizes ranging from 5 to 50 and solved the Max-Cut problem. The algorithm was executed for 1,000 iterations with a total of 250 initial conditions for each problem size. The ground state obtained from each run was used to calculate the corresponding cut values. If the ratio of the computed cut value to the optimal cut value was equal to or greater than 0.95, the solution was considered successful. We have also characterized the system under phase and coupling errors with standard deviations in the range of [0.0, 0.025, 0.05, 0.075, 0.1]. Simulation results for the success probability under varying phase and coupling errors are presented in Fig. \ref{fig:simulation}\textbf{a}-\textbf{b}, respectively. For the ideal case of $\sigma_{\phi} = 0$, the success probabilities remain above 0.8 across all problem sizes, approaching 1.0 for $N \leq 20$. However, as $\sigma_{\phi}$ increases from 0.05 to 0.1, the success probability drastically decreases from 0.8 to below 0.1, especially for problem sizes beyond $N = 30$. Similarly, our simulations show that coupling errors have a greater impact on success probability than phase errors (see Fig. \ref{fig:simulation}\textbf{b}). When coupling errors are below $\sigma_k = 0.025$, the success probability remains higher than 0.8. However, for $\sigma_k = 0.1$, the success probability drops significantly for $N > 20$. Next, we solved a M\"obius-ladder graph problem under varying phase and coupling errors using the simulator, with a problem size of $N = 20$. The success probability at $\alpha = 0.95$ was evaluated after a fixed number of 1,000 iterations. A two-dimensional plot of $\sigma_{\phi}$ and $\sigma_k$, shown in Fig. \ref{fig:simulation}\textbf{c}, illustrates that high success probability is maintained only at low error levels, thus any error in the phase value or coupling leads to poor solution quality. The evolution of the Hamiltonian for the ideal case of $\sigma_{\phi} = \sigma_k = 0$ is presented in Fig. \ref{fig:simulation}\textbf{d}. Our results are consistent with previous demonstrations \cite{Ouyang2024On-demandDetection, OUYANG2025100117}, where success probabilities approach 1 in fewer than 1,000 iterations. We further performed simulations for $N = 100$, with results shown in Fig. \ref{fig:simulation}\textbf{e}. A success probability of 0.8 was achieved over $2 \cdot 10^5$ iterations, corresponding to a time frame of approximately 20 $\mu$s. To achieve comparable performance to that reported in \cite{Fan:23}, we attained a success probability of 0.8 within 18 $\mu$s using an input modulation of 10 GHz.
In addition to success probability, the scalability of the photonic processor also depends on insertion losses or the precision required for computation. Our photonic architecture includes $(N+1)$ MZIs in the input stage, $2N$ MZIs in the matrix multiplication stage, and one additional MZI in the diagonal matrix stage before photodetection. As a result, the total insertion loss introduced by the MZI network is given by $(3N + 2)IL_{\text{MZI}}$, where $IL_{\text{MZI}}$ represents the insertion loss of a single MZI. The acceptable insertion loss of the processor is determined by the target bit precision, which is related to the signal-to-noise ratio (SNR) via the expression $N_{\text{bit}} = (\text{SNR} - 1.76)/6.02$. The insertion loss of the MZIs has been demonstrated to be as low as 0.05–0.1 dB. Using the value, we calculated the maximum achievable matrix size (N) for different bit precisions and insertion loss values \cite{9895610, Alexander2025AComputing}. Analog Ising machines typically require a bit precision between 6 and 8 bits \cite{Sevenants2025RequirementsVariables}. The results, presented in Table \ref{tab:max_matrix_size}, indicate that reducing the insertion loss of MZIs or relaxing the bit precision requirements can significantly improve scalability, enabling the implementation of larger problem sizes.

Furthermore, we evaluated the fidelity and MSE by configuring a random unitary matrix over 10,000 iterations. The measured fidelity is presented in Fig.~\ref{fig:stability}\textbf{a}, with a standard deviation of 0.0007 and a corresponding MSE value of 0.004. Although the photonic chip was placed within a temperature-controlled environment, we observed that even minor temperature fluctuations can introduce reflections or parasitic effects, leading to slight deviations in matrix fidelity over time. Since temperature variation and noise levels significantly impact system stability, we monitored the thermal environment using a thermistor and maintained operation at a constant temperature of \SI{25}{\celsius}. The measured system noise floor was \SI{-39.5}{\decibelm}, indicating excellent stability for Hamiltonian measurements. The recorded temperature fluctuations and noise measurements over the duration of the experiment are shown in Fig.~\ref{fig:stability}\textbf{b}–\textbf{c}. 

\begin{table}[h]
    \centering
    \caption{Maximum matrix size $N$ for different insertion losses and bit precisions.}
    \begin{tabular}{ccccc}
        \hline
        \textbf{MZI Insertion Loss (dB)} & \textbf{6 bits} & \textbf{7 bits} & \textbf{8 bits} \\
        0.45 & 14 & 10 & 5 \\
        0.10 & 68 & 48 & 27 \\
        0.05 & 136 & 96 & 56 \\
        \hline
    \end{tabular}
    \label{tab:max_matrix_size}
\end{table}

\subsection*{Latency and Energy Efficiency} 
Furthermore, we analyzed the latency and power consumption of the proposed photonic Ising architecture. In an Ising machine, once the eigenvector and eigenvalue matrices are programmed, they remain fixed; only the spin configuration requires updates. Therefore, the primary contributors to system latency are: (i) the time required for spin updates, (ii) the electronics executing the algorithm, and (iii) the forward and feedback latency within the system. Currently, all phase actuators rely on the thermo-optic effect, which is inherently limited to microsecond-scale speeds (MHz range) \cite{Siew2021}. However, latency can be significantly reduced by replacing the input vector array elements with high-speed modulators for spin encoding and updates, and by incorporating high-speed photodetectors \cite{10116484}. Additionally, using state-of-the-art field-programmable gate arrays (FPGAs), which operate at multi-GHz clock frequencies, can further reduce the overall latency to the nanosecond regime. Similarly, the throughput of the system can be estimated in terms of multiply-and-accumulate operations per second (MAC/s), calculated as $ \text{MAC/s} = 2N^2B $, where $N$ is the matrix dimension and $B$ is the bandwidth of the system. To estimate the total power consumption of the photonic Ising processor, we consider all contributing components, including the laser source, the $N$ input modulators, the thermo-optic phase shifters (used in both the splitter tree and matrix stages), the $N$ optical readouts, and the digital-to-analog converters:
\begin{equation}
    P_{\text{total}} = P_{\text{laser}} + NP_{\text{modulators}} + 2(N^{2} - 1)P_{\text{TO}} + NP_{\text{readout}} + NP_{\text{DAC}}.
\end{equation}
The power consumption of the thermo-optic elements was calculated by accounting for the entire optical architecture, including the splitter tree ($2(N-1)$ elements), the single-drive MZIs ($N(N-1)$) and the dual-drive MZIs ($\frac{N(N-1)}{2}$) used in the matrix multiplication. We assume the following power consumption values based on recent literature: 180 mW per digital-to-analog converter (DAC) \cite{Filipovich:22}, 480 mW per high-speed Mach–Zehnder modulator (MZM) \cite{10092945}, 32 mW (equivalent to 15 dBm) of input optical power with a laser wall-plug efficiency of 10\%, 1.3 mW per thermo-optic actuator \cite{Perez-Lopez2024}, and 48 mW per optical receiver \cite{Kim:21}. Using these parameters, we estimate both the total power consumption (in watts) and the computational efficiency (pJ/MAC). For an assumed operation speed of 5 GHz, we have estimated the computational speed, power consumption, and energy/operation for different matrix sizes and presented the results in Table \ref{tab:computational_speed}.

\begin{table}[h!]
\centering
\caption{Speed and power consumption characteristics of the programmable Ising machine}
\begin{tabular}{cccccc}
\hline
\textbf{N} & 4 & 8 & 16 & 32 & 64 \\
\textbf{Computational Speed (TMAC/s)} & 0.16 & 0.64 & 2.56 & 10.24 & 40.96 \\
\textbf{Power Consumption(W)} & 1.28 & 2.32 & 4.66 & 10.34 & 25.68 \\
\textbf{Energy/operation (pJ/MAC)} & 7.99 & 3.63 & 1.82 & 1.00 & 0.63 \\
\hline
\end{tabular}
\label{tab:computational_speed}
\end{table}

\subsection*{Stability, Operating Temperature and Noise effects} 
\begin{figure}[ht]
    \centering
    \includegraphics[width=1.0\textwidth]{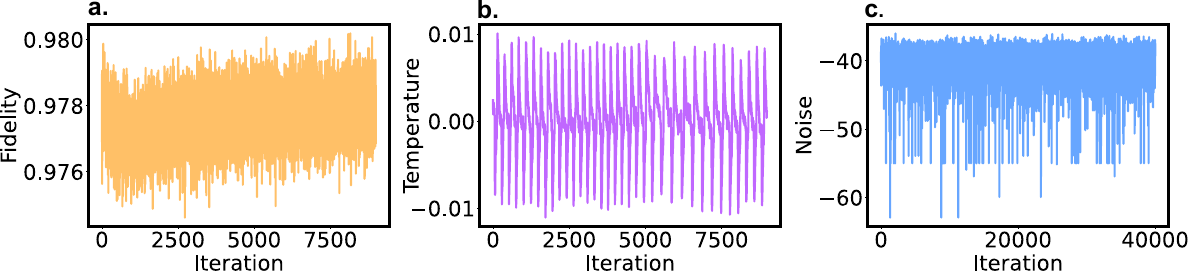} 
    \caption{(a) Stability Analysis via measurement of fidelity, (b) Temperature variation monitoring, (c) Impact of System noise during hamiltonian measurement}
    \label{fig:stability}
\end{figure}

The stability of the general-purpose processor was evaluated by configuring a random unitary matrix and measuring its fidelity and MSE with respect to the target matrix over 10,000 iterations. The fidelity results are presented in Fig.~\ref{fig:stability}\textbf{a}, showing a standard deviation of 0.0007, which corresponds to an MSE of 0.004. During the test, we also monitored temperature variations using a thermistor. The system was set to operate at \SI{25}{\celsius}, and maximum deviations of only \SI{0.01}{\celsius} were observed. Finally, the noise floor of the system was measured with the laser turned off, yielding an average photodetector noise level of \SI{-39.5}{\decibel}m. These results indicate that the system exhibits excellent stability for hamiltonian measurement. However, small temperature variations may impact the performance of the PUCs that must be configured in either the cross or bar state during matrix operations. Such variations can induce reflections or parasitic effects, which combined with the intrinsic noise of the integrated photodetectors lead to the slight deviations observed in matrix fidelity over time.

\section{Conclusions}
In conclusion, we have demonstrated a programmable photonic Ising machine based on a hexagonal mesh architecture that uses a simple phase encoding and intensity detection scheme. The programmable photonic circuit computes the Ising Hamiltonian through on-chip matrix-vector multiplications in the optical domain, while electronic components handle spin updates. For an efficient annealing algorithm, we employed a probabilistic update rule to iteratively minimize the energy of the system by probabilistically updating the spin configuration. Using the programmable photonic processor, we achieved MSE values of 0.007 and 0.016 for 4000 instances of \(3 \times 3\) and 6400 instances \(4 \times 4\) matrix multiplications, respectively. As a proof of concept, we solved a three-node interaction problem across 25 different coupling strengths, demonstrating reconfigurability and convergence to the ground-state solution in fewer than 50 iterations, with success probabilities of unity in most cases. Additionally, we solved a four-node Max-Cut problem using five randomly generated coupling matrices, successfully identifying the optimal solution in fewer than 100 iterations, achieving near-perfect fidelity between ideal and experimental cut values with a success probability exceeding 0.998 for all cases. Furthermore, to assess the scalability of our architecture, we simulated its performance for larger problem sizes (\(N \geq 50\)), demonstrating feasibility with current state-of-the-art silicon-on-insulator (SOI) technology. We also evaluated the impact of phase noise and fabrication errors, demonstrating that our system maintains a success probability above 0.8 for \(N = 50\), given error levels within current fabrication tolerances. Compared to alternative platforms such as the D-Wave quantum system, our photonic Ising machine offers several advantages, including room-temperature operation, overcoming low coherence, and improved energy efficiency. Furthermore, performance comparisons with free-space optical approaches indicate competitive results in terms of computational speed, energy consumption, system stability, latency, and temperature robustness. Our results demonstrate a simple yet powerful on-chip reconfigurable photonic architecture, establishing the photonic Ising machine as a promising platform for solving large-scale combinatorial optimization problems.

\section{Methods}
\subsection*{Eigen decomposition based Ising solver}\label{section:PEIDIA}
The energy landscape of the optimization problem is defined by the Hamiltonian of the Ising model, expressed as:
\begin{equation}
    H(\sigma) = -\frac{1}{2} \sigma^T J \sigma
    \label{eq:hamiltonian}
\end{equation}

Where $J$ represents the matrix encoding the coupling strengths between the spins, and $\sigma$ is the spin configuration vector that only takes values $\pm 1$. 

In the Ising model, the matrix $J$ is symmetric and thus, we can take advantage of the properties of this type of matrices.  A symmetric matrix can easily be diagonalized with its eigenbasis such that one can write back $ J $ as $ J = Q^T \Lambda Q $. Here, $ Q $ is an orthogonal matrix whose columns are the eigenvectors of $ J $, and $ \Lambda $ is a diagonal matrix containing the eigenvalues of $ J $. For further simplification, $\Lambda$ can be decomposed into the multiplication of two diagonal matrices $\Lambda = \sqrt{D}\sqrt{D}$. 

Substituting this decomposition into (\ref{eq:hamiltonian}), we get:
\begin{equation}
    H(\sigma) = -\frac{1}{2} \sigma^T Q^T \sqrt{D}\sqrt{D} Q \sigma = -\frac{1}{2}(\sqrt{D}Q\sigma)^T(\sqrt{D}Q\sigma)
    \label{eq:H_decomp}
\end{equation}

We can encode the vector-matrix multiplication $\sqrt{D}Q\sigma$ in the optical domain by normalizing its elements beforehand. After photodetection, the multiplication returns:
\begin{equation}
    (\sqrt{D}Q\sigma)\cdot(\sqrt{D}Q\sigma)^*
    \label{eq:H_decomp_2}
\end{equation}
where $^*$ represents the complex conjugate and $\cdot$ represents element-wise multiplication.
The only difference between (\ref{eq:H_decomp}) and (\ref{eq:H_decomp_2}) apart from the scaling terms is the conjugate of the results. This operation during photodetection results in all-positive photocurrents. We know that the diagonal matrix includes real and imaginary numbers $\sqrt{D} = diag(i\sqrt{\lambda_{1}}, i\sqrt{\lambda_{2}}, ..., i\sqrt{\lambda{n}})$ that during the multiplication by the transpose operators in (\ref{eq:H_decomp}) become negative elements. As a consequence, if we change the sign of the photocurrents belonging to negative eigenvalues, we can calculate the energy of the system with only direct photodetection as follows:

\begin{equation}
    H(\sigma) = \frac{1}{2} \left( \sum_{\lambda_i < 0} I_i - \sum_{\lambda_i > 0} I_i \right)
    \label{eq:H_photodetection}
\end{equation}
where $ I_i $ are the optical intensities corresponding to the eigenvalues $ \lambda_i $ of $ J $.

Following the Hamiltonian calculation, the annealing algorithm iteratively updates the spin configuration to locate the global energy minimum. The annealing process begins with the generation of an initial state, where the spin vector $\sigma^{(0)}$ is initialized randomly. At every step, a subset of spins is flipped at random to produce a candidate configuration $\sigma^{(n+1)}$. The acceptance of the new configuration is determined using the following criterion: if the change in the Hamiltonian $\Delta H$ is less than or equal to zero, the configuration is accepted; otherwise, it is accepted with a probability $P(\text{accept}) = \exp(-\Delta H / T)$. The temperature $T$ is gradually reduced according to a predefined cooling schedule, allowing broad exploration at high temperatures and finer refinements as the temperature decreases. The process stops when the maximum number of iterations is reached, and the configuration with the minimum observed Hamiltonian is selected as the solution. This process of Hamiltonian calculation and annealing algorithm can be translated into a photonic-electronic feedback-based hardware as the one depicted in Fig. \ref{fig:conceptual}\textbf{a}. First, the coupling matrix $J$ is decomposed electronically, and the resulting matrices are encoded in the optical processor where they remain fixed during the whole operation. Then, the algorithm is initialized with a spin configuration $\sigma^{(0)}$ that is also encoded in the optical system and the vector-matrix multiplication is measured. Photocurrents are used for Hamiltonian calculation, and a new spin configuration is obtained through the annealing algorithm. The feedback loop continues until the maximum number of iterations is reached.

\subsection*{Smartlight Processor}
\begin{figure}[h]
    \centering
    \includegraphics[width=0.9\textwidth]{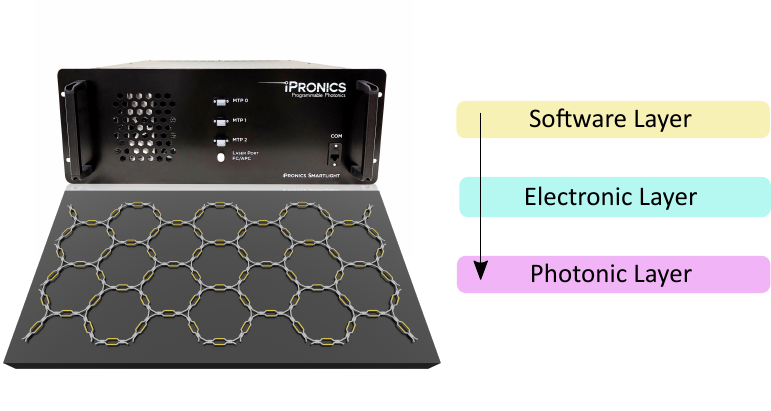} 
    \caption{\textbf{a.} Smartlight processor from IPronics. It combines a hexagonal waveguide mesh with an electronic and software layer. (b) Programmable unit cell (PUC) of the hexagonal mesh. It consists of two internal phase shifters $\theta_1$ and $\theta_2$. }
    \label{fig:ipronics}
\end{figure}

The Smartlight processor, see Fig. \ref{fig:ipronics}, includes 17 hexagonal cells, for a total of 72 MZIs each with an insertion loss of 0.45 dB and a power consumption of 1.3 mW/$\pi$. The system supports light coupling through 28 optical ports, which are accessible via a fiber array that introduces a 3-dB insertion loss per facet. Each optical port is equipped with an on-chip photodetector for direct power measurement, eliminating the need for external monitoring devices. Additionally, the processor includes integrated electronics and a comprehensive software suite that enable precise configuration and control of the PUCs. Further technical specifications and details on the device can be found in \cite{Perez-Lopez2024}.
Simulations were performed on an Intel Core i5-10400H 2.6 GHz CPU with 16 GB of RAM.

\begin{backmatter}
\bmsection{Funding}
This work was supported by the European Research Council (ERC) Advanced Grant program under grant Agreement No. 101097092 (ANBIT), the ERC Starting Grant program under grant Agreement No. 101076175 (LS-Photonics Project), and the COMCUANTICA/005 and COMCUANTICA/006 grants, funded by the European Union through NextGenerationEU (PRTR-C17) with the support of the Spanish Ministry of Science and Innovation and the Generalitat Valenciana.

\bmsection{Acknowledgment}

\bmsection{Disclosures}
The authors declare no conflicts of interest.

\bmsection{Data Availability Statement}
Data underlying the results presented in this paper are not publicly available at this time but may be obtained from the authors upon reasonable request.

\end{backmatter}

\bibliography{sample, references}

\end{document}